\documentclass[superscriptaddress,nofootinbib,nobibnotes,amsmath,amssymb,aip,preprint]{revtex4-2}
\usepackage{graphicx}
\usepackage{dcolumn}
\usepackage{bm}
\usepackage[T1]{fontenc}
\usepackage{mathptmx}
\usepackage{color}
\usepackage{natbib}
\usepackage{mathtools}

\begin{document}

\preprint{APS/123-QED}

\title{Nuclear-induced dephasing and signatures of hyperfine effects in isotopically purified $^{13}$C graphene}

\author{Vincent Strenzke}
\email[]{Correspondence: vstrenzk@physnet.uni-hamburg.de}
\affiliation{Center for Hybrid Nanostructures, Universit\"at Hamburg, Luruper Chaussee 149, 22761 Hamburg}
\author{Jana M. Meyer}
\affiliation{Center for Hybrid Nanostructures, Universit\"at Hamburg, Luruper Chaussee 149, 22761 Hamburg}
\author{Isabell Grandt-Ionita}
\affiliation{Center for Hybrid Nanostructures, Universit\"at Hamburg, Luruper Chaussee 149, 22761 Hamburg}
\author{Marta Prada}
\affiliation{I. Institute for Theoretical Physics, Universit\"at  Hamburg, Luruper Chaussee 149, 22761 Hamburg}
\author{Hyun-Seok Kim}
\affiliation{Dongguk University, Seoul}
\author{Martin Heilmann}
\email[]{Present address: IMEC, Leuven/Belgium}
\affiliation{Paul-Drude-Institut f\"ur Festk\"orperelektronik, Leibniz-Institut im Forschungsverbund Berlin e.V., Berlin}
\author{Joao Marcelo J. Lopes}
\affiliation{Paul-Drude-Institut f\"ur Festk\"orperelektronik, Leibniz-Institut im Forschungsverbund Berlin e.V., Berlin}
\author{Lars Tiemann}
\affiliation{Center for Hybrid Nanostructures, Universit\"at Hamburg, Luruper Chaussee 149, 22761 Hamburg}
\author{Robert H. Blick}
\affiliation{Center for Hybrid Nanostructures, Universit\"at Hamburg, Luruper Chaussee 149, 22761 Hamburg}

\date{\today}
\newpage 

\begin{abstract}
The  hyperfine interaction between the spins of electrons and nuclei is both a blessing and a curse. It can provide a wealth of information when used as an experimental probing technique but it can also be destructive when it acts as a dephasive \textcolor{black}{perturbation} on the electronic system. In this work, we fabricated large scale single and multilayer isotopically-purified $^{13}$C graphene Hall bars to search for interaction effects between the nuclear magnetic moments and the electronic system. We find signatures of nuclei with a spin in the analysis of the weak localization phenomenon that shows a significant dichotomy in the scattering times of monolayer $^{12}$C and $^{13}$C graphene close the Dirac point. Microwave-induced electron spin flips were exploited to transfer momentum to the nuclei and build-up a nuclear field. \textcolor{black}{The presence of a very weak nuclear field is encoded in a modulation of the electron Zeeman energy which shifts the energy for resonant absorption and reduces the $g$-factor.} 
\end{abstract}

\maketitle
\noindent  

\section{INTRODUCTION}

Phase coherence is an integral element of quantum physics and imperative for the observation of quantum effects in two-dimensional carrier systems. Its susceptibility to magnetic/nonmagnetic \textcolor{black}{perturbations} and/or charged/uncharged defects that cause dephasing is limiting the distances for coherent electron transport in practical applications. In applied quantum physics, various sources lead to dephasing that  depend on the host material itself, its growth and in many cases also on the preparation methods. Dephasing of localized electron spin states in III/V semiconductor nanostructures, for example, is a caveat for the electrical control of isolated solid state qubits. The decay of the stored spin quantum information is fueled by randomly fluctuating nuclear magnetic moments in materials that possess a nonzero nuclear spin $I$. 

The abundance of nuclei with a spin depends on the host material for the conducting electrons or holes. AlGaAs/GaAs heterostructures, which are the foundation for high mobility two-dimensional electron systems that also led to the discovery of the fractional quantum Hall effect for example, are built up entirely of isotopes with $I$ > 0. Silicon consists by almost 95$\%$ of $^{28}$Si isotopes that have no nuclear spin. Graphene, as exfoliated from graphite\cite{Novoselov2005}, is comprised by 99$\%$ of $^{12}$C isotopes with $I$ = 0 and by only 1$\%$ $^{13}$C isotopes with $I$ = 1/2. The presence of nuclei with a nuclear spin $I$ > 0 gives rise to the hyperfine interaction. The hyperfine interaction depends on the density of the electron spin at the site of the nucleus and a dipole interaction between the magnetic moments of electron and nuclei. In naturally \textcolor{black}{occurring} $^{12}$C graphene, the hyperfine interaction is expected to be small\cite{Yazyev2008, Fischer2009, Wojtaszek2014} due to the highly diluted nuclear magnetic moments and the $\pi$-orbitals that have a vanishing probability density at the site of the nuclei. All the same, it was shown that magnetic moments in general are a significant source of (orbital) dephasing in graphene\cite{Lundeberg2013}. Graphene with an artificially maximized number of nuclear spins is expected to leave an imprint on electron transport and --mediated by the hyperfine interaction-- also on the electron spin dynamics.

In this study, we will search for imprints of the nuclear magnetic moments and the hyperfine interaction on carrier transport and electron spin dynamics in graphene. In a careful study, Wojtaszek \textit{et al.} \cite{Wojtaszek2014} had searched for signatures of the hyperfine interaction using a $^{13}$C-graphene spin-valve devices in Hanle precession measurements. The authors observed only a negligible effect of the hyperfine interaction in spin transport in their experiments at 4.2 K and low magnetic fields $B$ < 0.2 T. Here, we revisit this very important issue as the $p$-orbital character of carriers in graphene is shared by the whole family of van der Waals materials. We use isotopically purified large scale single and multilayer $^{13}$C graphene synthesized by chemical vapor deposition (CVD) and molecular beam epitaxy (MBE), respectively, to understand how the nuclear spin bath affects coherence and electron spin dynamics. 

From the presence of the weak localization (WL) phenomenon in $^{12}$C and $^{13}$C single layer graphene, we deduce the phase-breaking, intralayer and interlayer scattering times. Single layer $^{13}$C graphene near the Dirac point (DP) has significantly lower scattering times. Contrary to earlier reports\cite{Wojtaszek2014}, we attribute this behavior to the scattering on nuclear magnetic moments that is mediated by the hyperfine interaction at low carrier densities, a regime where screening is weak. Microwave-activated magneto-transport measurements on multilayer $^{13}$C graphene were used to induce electron spin flips. The hyperfine interaction should facilitate a momentum transfer to the nuclei and the build-up of a nuclear polarization. We find weak signatures of a dynamical polarization of the nuclear bath \textcolor{black}{as small shifts in the energy for resonant absorption and changes in the deduced electron $g$-factor.}

\section{SAMPLE FABRICATION AND EXPERIMENTAL METHODS\label{sec:sample}}

\begin{figure}[!]
    \includegraphics[width=0.5\textwidth, angle=0]{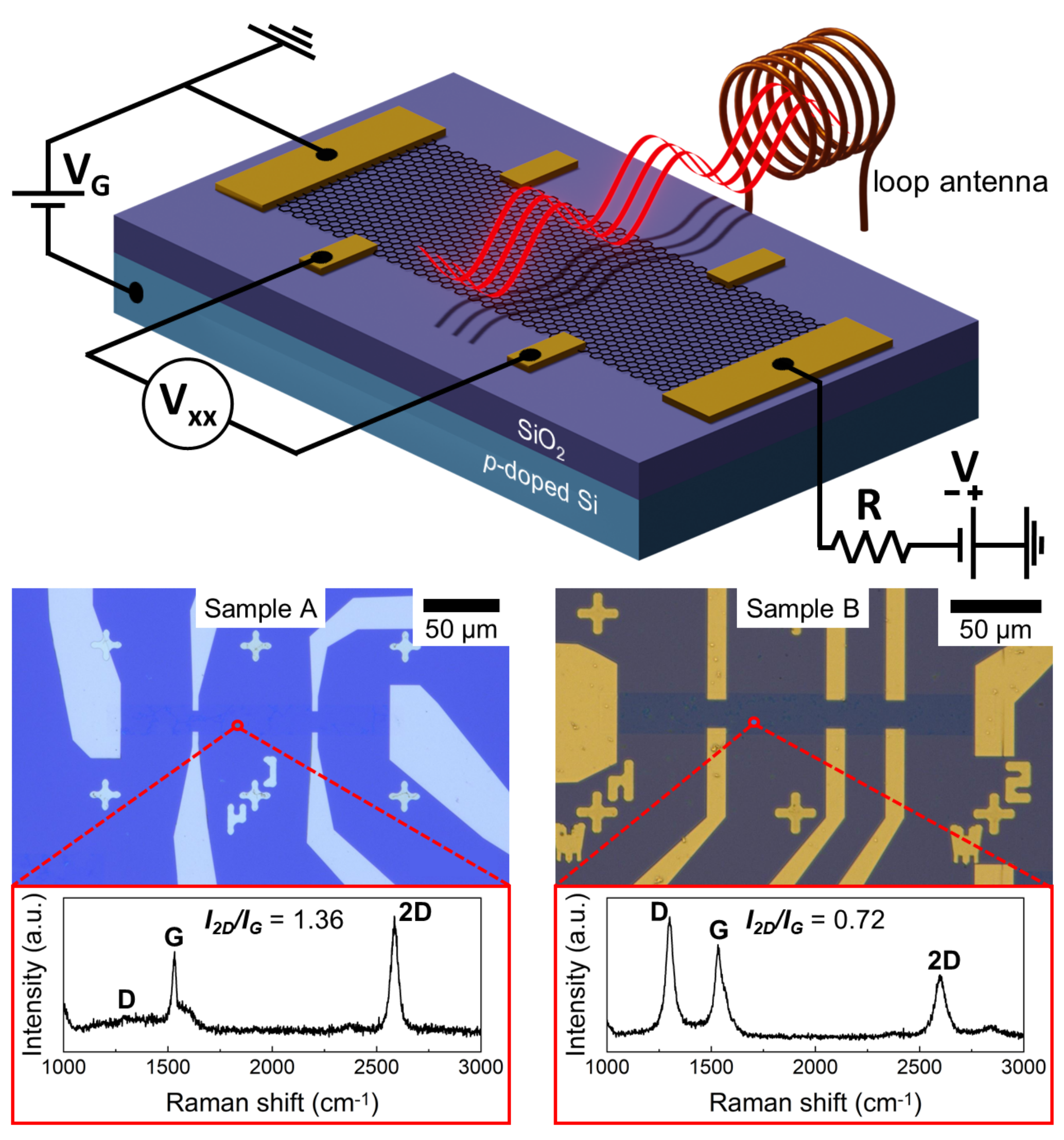}  
   \caption{\textcolor{black}{(top panel)} Schematics of the Hall bar structure and measurement setup.  A constant low frequency alternating current is passed between a source and a drain contacts at opposite ends of the Hall bar and the longitudinal voltage, $V_\mathrm{xx}$, is monitored at one edge. A loop antenna exposes the sample to microwave radiation (only for section \ref{sec:esr}). \textcolor{black}{(center panels) A microscopic image of sample A and of sample B . The analysis of corresponding Raman spectra (bottom panels) in combination with AFM measurements indicate that sample A is single layer $^{13}$C graphene and sample B few-layer $^{13}$C graphene with 2 to 4 layers.}}
  \label{fig1}
\end{figure}

Monolayer CVD $^{13}$C graphene was synthesized using an ethylene precursor with isotopically purified $^{13}$C atoms to \textcolor{black}{deposit} large scale graphene on a copper foil. Polycrystalline foils were also used as templates for MBE growth of multilayer graphene. Here, growth is performed in ultra-high-vacuum utilizing an electron-gun evaporator loaded with an isotopically enriched $^{13}$C powder. Two devices using monolayer $^{13}$C-graphene (sample A) and few-layer $^{13}$C-graphene (sample B) were fabricated. In a wet transfer process, the graphene is placed on a 300 nm layer of SiO$_2$ on top of a degenerately $p$-doped Si substrate after removing the copper in an iron nitrate solution followed by multiple cleaning steps in deionized water.\cite{Lyon2019} In a thermal annealing step at 350$^\circ$C for 2 hours and 200$^\circ$C for 12 hours adhesion to the substrate is ensured and chemical residues remaining on the surface are removed. \cite{Shautsova2016, DeFazio2019, Kim2015, Yankowitz2012} To fabricate the 180 $\times$ 22 $\mu$m$^2$ Hall bar structures, the graphene is patterned by photolithography and O$_2$ - plasma etching, followed by physical \textcolor{black}{vapor} deposition of Ti/Au (7 nm/70 nm) metal contacts to the graphene after electron beam lithography.

The insets in Figs. \ref{fig1} b) and c) show Raman characterization spectra for samples A and B. Raman spectroscopy excites vibrational modes that are inversely proportional to the square root of the atomic mass, making it sensitive to the mass constituents of carbon\cite{Rodriguez-Nieva2012,delCorro2013}. \textcolor{black}{The downward shift of the Raman peaks towards lower wavenumbers by a factor of $\sqrt{13/12}$ is characteristic for isotopically purified $^{13}$C-graphene due to the mass difference between the $^{12}$C and $^{13}$C isotopes.\cite{Bernard2012} The 2D peaks appear around 2585\,cm$^{-1}$ for sample A and 2600\,cm$^{-1}$ for sample B. For both samples, the G and D bands are around 1530 and 1300\,cm$^{-1}$, respectively. The observed shifts are similar to those of previous Raman studies on $^{13}$C-graphene.\cite{Wojtaszek2014,Carvalho2015} The number of graphene layers can be deduced from the intensity ratio $I_\mathrm{2D}$/$I_\mathrm{G}$ of the 2D and G peaks and additional atomic force microscopy (AFM) measurements. Our characterization that is based on a combination of Raman spectroscopy and AFM measurements} confirms that sample A consists of single layer $^{13}$C-graphene, whereas sample B is $^{13}$C-graphene of two to four layers. In contrast to sample A, the Raman spectrum of sample B shows a significantly larger D peak, which is associated with a higher defect density.

The samples were cooled down in vacuum to liquid helium temperatures to study the dephasing by nuclear magnetic moments. A standard low-frequency lock-in technique is used to detect the longitudinal voltage, $V_{\mathrm{xx}}$, under perpendicular magnetic fields that acts as a probe. Only for the microwave-activated transport that is discussed in section \ref{sec:esr}, we make use of a loop antenna positioned close to the sample, as illustrated in Fig. \ref{fig1}. All samples are equipped with a back-gate to control the charge carrier concentration and carrier type.

Due to unintentional doping, the samples are intrinsically hole-doped. A back-gate voltage of 15 V (80 V) for sample A (sample B) is required to lower the density and shift the Fermi level to the charge neutrality point (CNP). For sample A, the maximum charge carrier mobility $\mu$ deduced from the measurement of the Hall voltage is 552 cm$^2$/Vs while the residual charge carrier density at the CNP $n^*$ is of the order of 10$^{12}$ cm$^{-2}$. All measurements on sample B were performed with \textcolor{black}{a four-terminal configuration but without Hall probes.}

We note in passing that the carrier mobility in large-scale graphene synthesized on polycrystalline foils can not approximate that of small flakes of exfoliated graphene. Scattering is dominated by grain boundaries, wrinkles, folds and defects as well as disorder introduced in the transfer process\cite{Kim2015,Petrone2012,Zhu2012}. For us, mobility is not the most important figure of merit, though. Synthesized graphene allows the fabrication of ultra-large scale samples, better coupling and larger signal intensities.


\section{QUANTUM INTERFERENCE\label{sec:wal}}

We use the weak localization (WL) phenomenon to probe the presence of nuclei with a magnetic moment. WL is a quantum correction to the Drude resistance in the absence of a magnetic field. It arises from the constructive interference of two time-reversed electronic paths enclosing a loop in the clockwise and anti-clockwise direction. If the phase is preserved over both paths, constructive interference at the point of origin leads to an increased possibility of backscattering.

Application of an external magnetic field \textit{\textbf{B}} breaks the time reversal symmetry and cancels constructive interference by introducing a phase difference between the two paths. \textcolor{black}{WL} thus diminishes monotonically with \textit{\textbf{B}}. The associated theory and mathematical description of this phenomenon was developed by McCann \textit{et al.} [\onlinecite{McCann2006}] and can be expressed as

\begin{equation}
\begin{multlined}
\Delta\rho=\rho(B)-\rho(0)= \\
-\frac{e^{2} \rho^{2}}{\pi h}\left[F\left(\frac{B}{B_{\phi}}\right)-F\left(\frac{B}{B_{\phi}+2 B_{i}}\right)-2 F\left(\frac{B}{B_{\phi}+B_{i}+B_{*}}\right)\right]
\end{multlined}
\label{eqn:mccann}
\end{equation}

with $F(z)=\ln z+\Psi\left(\frac{1}{2}+\frac{1}{z}\right)$ and 
$B_{\phi, i, *}=\frac{\hbar}{4 D e} \tau_{\phi, i, *}^{-1}$ which include the diffusion coefficient $D = \frac{1}{2}v_{\text{F}} l_{\text{mfp}}$ and the mean free path for elastic scattering  $l_{\text{mfp}}$. 

\begin{figure}[!]
   \includegraphics[width=0.45\textwidth, angle=0]{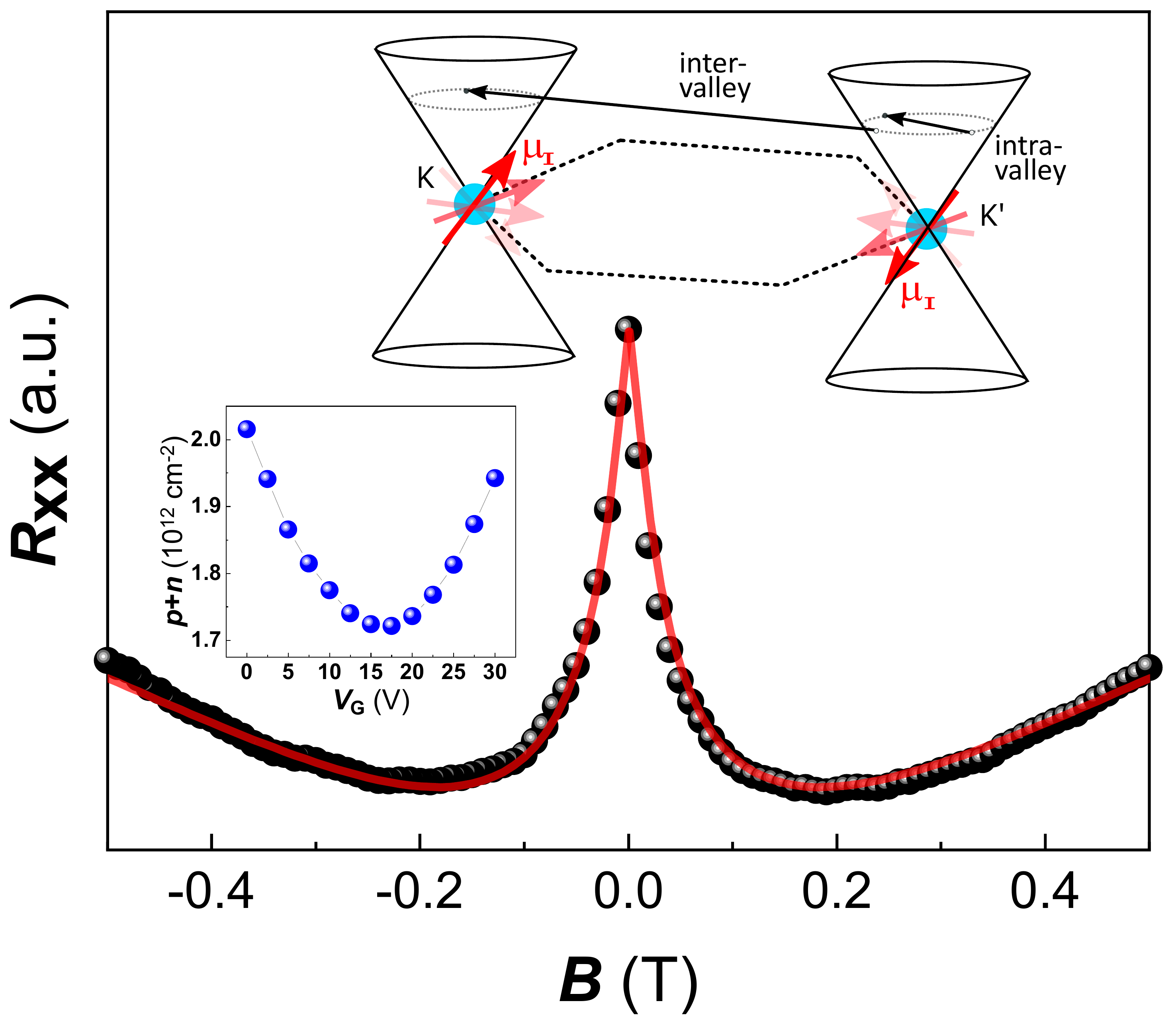}
   \caption{\textcolor{black}{Exemplary} weak localization measurement (black dots) of sample A at 1.5 K. The red solid line shows the fit using Eq. (\ref{eqn:mccann}). 
The upper inset illustrates two Dirac cones in the honeycomb lattice with  intra- and intervalley scattering events. The circles at the Dirac points represent  $^{13}$C nuclei with their spin magnetic moments (red arrows) $\mu_\mathrm{I}$ around $B \approx$ 0 T.  The lower left inset shows the density calibration according to Eqs. \ref{eqn:2CMrhoxx0} and \ref{eqn:2CMn-p}.}
  \label{fig2}
\end{figure}

The main panel of Fig. \ref{fig2} shows an exemplary magneto-transport trace of the WL phenomenon measured in sample A at 1.7 Kelvin (black circles). The red solid line is a fit using equation (\ref{eqn:mccann}) that yields the scattering times $\tau_{\phi,i,*}$ for phase coherence ($\phi$), intervalley (i) and intravalley (*) scattering. We will discuss the physical significance of these time scales below along with our experimental data.

We performed magneto-resistance measurements at different gate voltages and fitted the resulting WL peaks. The determination of the charge carrier density that requires a careful evaluation. Near the \textcolor{black}{DP}, the carrier density is generally not zero at finite temperatures and potential fluctuations induce electron-hole puddles. Hence, both electrons and holes close to the CNP contribute to transport. A simple one-carrier capacitive model can not capture this regime close to the CNP and it is thus more reasonable to rely on a two-carrier transport model as presented by Hilke \textit{et al.} [\onlinecite{Hilke2014}].

At low magnetic fields, shy of the WL regime but well before magneto-resistive effects kick in (\textsl{i.e.}, $\omega_{\text{c}}\tau\ll1$), the Drude model can still be employed to approximate the transport. By assuming a constant mobility $\mu=\mu_{\text{e}}=\mu_{\text{h}}$ and a total resistivity given by the contribution of each carrier type $\rho_{\text{tot}}=(\sigma_{\text{e}}+\sigma_{\text{h}})^{-1}$, the resistivity at $\bm{B}=0$~T (without WL) reads 

\begin{equation}
\label{eqn:2CMrhoxx0}
\rho_{\text{xx}}(0) \simeq \frac{1}{e \mu(p+n)},
\end{equation}

where \textit{n} (\textit{p}) are the individual densities for electrons (holes).
After extraction of the total carrier concentration $p+n$ from Eq. (\ref{eqn:2CMrhoxx0}), it is possible to calculate the net carrier charge $n-p$ using the slope of the Hall resistance at small magnetic fields, $\frac{\Delta\rho_{\text{xy}}}{\Delta B}\big(B \rightarrow 0\big)$,

\begin{equation}
\label{eqn:2CMn-p}
  n-p \simeq e(p+n)^2 \cdot \frac{\Delta\rho_{\text{xy}}}{\Delta B}.
\end{equation}

For gate voltages that shift the Fermi energy far away from the CNP, $n$ is extracted directly from the Hall resistivity. Close to the CNP, however, the charge carrier density is determined by Eq. (\ref{eqn:2CMn-p}). Derivation and details on the calculation of $n^{*}$ can be found in the work of Adam \textit{et al.} [\onlinecite{Adam2007}].
The lower inset in Fig. \ref{fig2} shows the corrected carrier concentrations as a function of the gate voltage. 

\begin{figure}[!]
   \includegraphics[width=0.45\textwidth, angle=0]{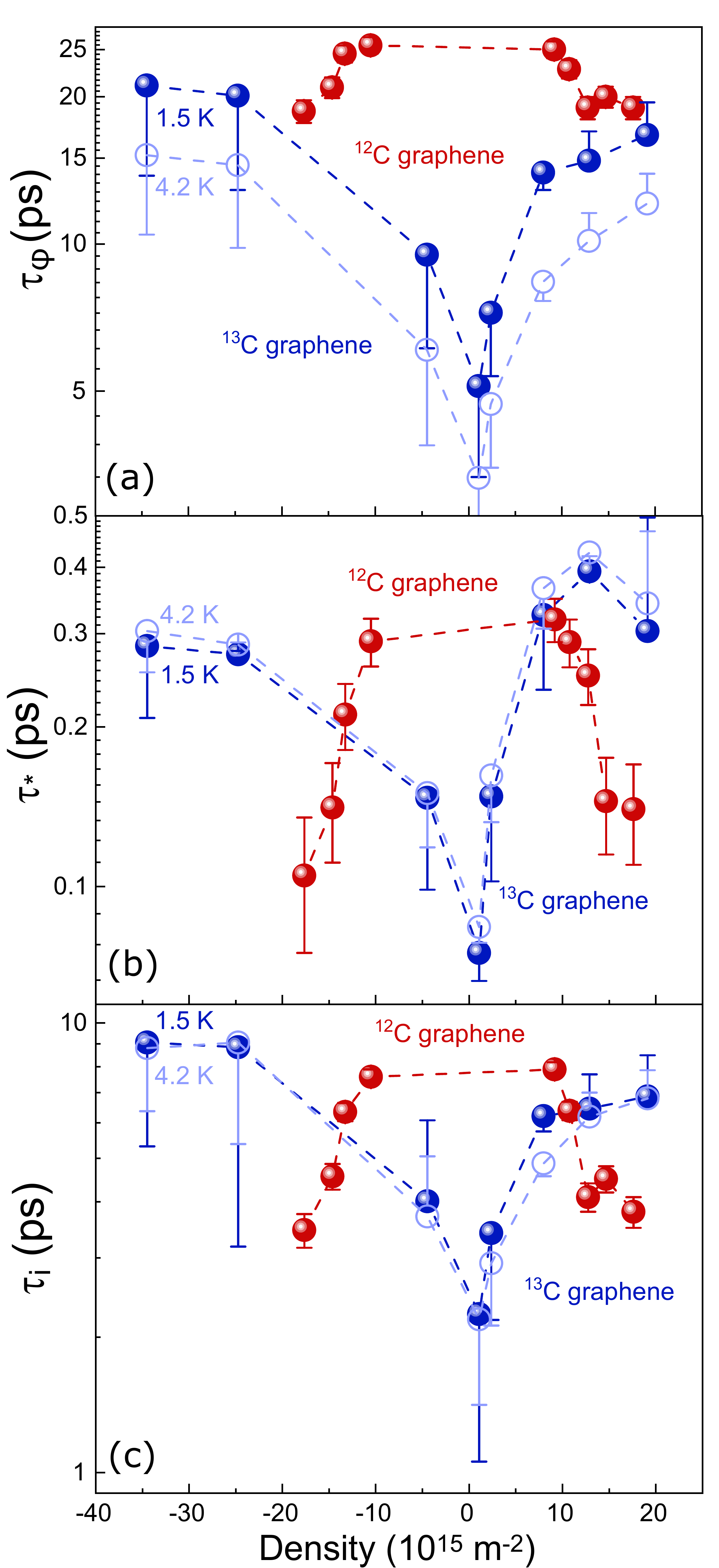}
   \caption{(a) phase-breaking scattering time $\tau_\phi$, (b) intravalley scattering time $\tau_*$ and (c) intervalley scattering time $\tau_i$ measured on sample A (single layer $^{13}$C-graphene, blue dots) and a reference sample of $^{12}$C-graphene (red dots). \textcolor{black}{The measurements were performed at 1.5 K (closed dots) and 4.2 K (open dots).} }
  \label{fig3}
\end{figure}

In Figs. \ref{fig3} (a)-(c), we now plot the \textcolor{black}{extracted} phase coherence time $\tau_{\phi}$, the intervalley scattering time $\tau_{i}$ and intravalley scattering time $\tau_{*}$ for sample A (blue circles) \textcolor{black}{and a reference sample of $^{12}$C CVD graphene with a carrier mobility of 1650 cm$^2$/Vs (red dots)} 
as a function of the calibrated densities.
\textcolor{black}{We note that the spinless model by McCann {\it et al.}
[\onlinecite{McCann2006}] can be justified within the quasi-degenerated spin model for the $^{13}$C sample. In order to have a direct comparison between both samples within a compact model, we avoid introducing additional free 
parameters that account for nuclei enhanced spin scattering terms. As a result, 
we expect an anomaly in the deducted scattering times, as hyperfine coupling 
enhances both elastic and inelastic processes \cite{Palyi2009}. }

\textcolor{black}{The most striking feature in these plots is the pronounced drop of all scattering times in the $^{13}$C sample when approaching the CNP. 
We can interpret this effect by noting that near CNP the magnetic moment density due to electron spin and nuclei spin becomes comparable: Although the nuclear magnetic moment $\mu_\mathrm{13C}$ of $^{13}$C nuclei is much smaller than the Bohr magneton
$\mu_\mathrm{B}$ of the electrons due to the higher nuclear mass, i.e., $\mu_\mathrm{13C}$ = $\mu_\mathrm{B}/2600$, the nuclear density $N$ is much larger than the typical carrier concentrations $n$ in graphene \cite{Wojtaszek2014}, specially near CNP.  
Moreover, at low carrier concentrations, i.e., close to the DP, a high concentration of carriers become locally trapped while at the same time carrier screening of the nuclear moments is weak. It is known that locally trapped carriers are susceptible to fast spin exchange \cite{Paget1977, Paget1981}. We thus hypothesize that similar effects -- mediated by the hyperfine interaction and the large number of weakly screened nuclei -- is involved in the suppression of all scattering times.}

\textcolor{black}{The dephasing rate $\tau_{\phi}^{-1}$ for $^{13}$C shows a similar behavior at 1.5 K (closed blue dots) and 4.2 K (open blue dots), except for an overall shift. Considering the dephasing rate as linear with temperature, $\tau_\phi^{-1} = \tau_s^{-1} + AT$ \cite{Long2014, LaraAvila2011}, we would obtain a faster spin relaxation rate $\tau_s^{-1}$ as the carrier concentration is close to the CNP. Moreover, $\tau_s^{-1}$ for the $^{13}$C sample is roughly one order of magnitude faster than  for $^{12}$C, which is in line with the arguments presented above.}

\textcolor{black}{We can further scrutinize the regular scattering mechanism [visualized in the upper inset of Fig. \ref{fig2}] that enhance or subdue WL to interpret the dichotomy between the $^{12}$C and $^{13}$C graphene scattering times near the CNP. First, we remind the reader of the character of the nuclear spins as: 1. (time-dependent) randomly fluctuating spin moments that are linked to the electron spins via the hyperfine interaction, and 2. atomically sharp impurities\cite{Palyi2009}. How can we reconcile these properties with the standard WL model?} 

\textcolor{black}{WL} originates from interference of two paths \textcolor{black}{which makes it} sensitive to inelastic phase-breaking scattering parameterized by $\tau_{\phi}$ and its competition with intervalley and intravalley scattering. Phase-breaking scattering is generally associated with inelastic phonon scattering, where the scattering target changes with time. There are, however, also elastic scattering events that can contribute to phase-breaking, such as spin-flip events. \textcolor{black}{Since nuclear moments are coupled to the electron spins via the hyperfine interaction, randomly fluctuating nuclear moments can transfer spin momentum to the electrons, which represents phase-breaking scattering.} 

Int\underline{er}valley elastic scattering ($\tau_{i}$) is promoted by sharp defect or edges that flip the direction of the carriers. We may expect this to be dominant near the DP, where the potential landscape consists of puddles of electron and holes. Intervalley scattering mixes two valleys that have opposite chirality and can thus restore WL. \textcolor{black}{As nuclear spins can be considered atomically sharp defects, it is feasible that their existence will enhance intervalley scattering.} 

Int\underline{ra}valley elastic scattering ($\tau_{*}$) originates from scattering on defects with sizes comparable to interatomic distances and the trigonal warping effect, i.e., a trigonal deformation of the Dirac cone. Trigonal warping will appear as we shift away from the DP. Although intravalley scattering is elastic, it still breaks the chirality within each of the two graphene valleys in $k$-space. Owing to the relative Berry phase of $\pi$, the backscattering interference is destructive at $B$ = 0. Intravalley scattering is thus a dephasive process that competes with the constructive WL interference. \textcolor{black}{Nuclear spins exist on interatomic distances on all atomic sites and may contribute to intravalley scattering.}

\textcolor{black}{Various defects exist that influence intra-, inter- and phase-breaking scattering regardless of the isotope.} Graphene synthesized by \textcolor{black}{CVD} is comprised of grains that form during growth when the nucleation centers fuse into a continuous layer and can exhibit additional lattice defects such as vacancies, line defects and folds\cite{Munoz2013,Zhang2013}. Vacancies in the crystal lattice are associated with localized states that possess magnetic moments\cite{Kozikov2012, Hsieh2021} and induce spin scattering. 
\textcolor{black}{The same types of defects will be present in both the $^{12}$C CVD graphene and $^{13}$C CVD graphene but their concentration differs such that the mobility of the $^{13}$C-graphene is smaller by a factor of 3 as compared to the $^{12}$C reference sample. We emphasize that the lower mobility alone can not account for the \textit{opposite trends} of the scattering times for $^{13}$C- and $^{12}$C-graphene graphene near the CNP.}

\textcolor{black}{The experimentally observed suppression of $\tau_{\phi}$ implies fast electron spin diffusion by momentum transfer between the electrons and nuclei.} This type of momentum transfer is a precursor for dynamic nuclear polarization (DNP). 
In the following section, we try to exploit this method to generate a sizable nuclear field via electron spin resonance.

\section{ELECTRON SPIN ACTIVATION AND HYPERFINE INTERACTION\label{sec:esr}}

In GaAs two-dimensional electron systems, the hyperfine interaction that is parameterized by the hyperfine constant $A_\mathrm{hf}$ (which in its most general form is a tensor) is three orders of magnitude larger than in graphene as a result of the $s$-orbital character of conduction band electrons\cite{Paget1977}. The large hyperfine interaction in GaAs enables the generation and detection of nuclear magnetic polarizations at low temperatures.
A nuclear polarization is tantamount to a nuclear magnetic field that modifies the electronic Zeeman splitting. The observation of an \textit{Overhauser shift} of the Landau level filling factor $\nu$ = 2/3 spin transition \cite{Kronmueller1998} is a prominent example. A strong  nuclear polarization out of thermal equilibrium can be achieved by means of DNP. DNP pumps the nuclear level polarization by exploiting momentum transfer when an electron flips its spin and transfers momentum to the nuclei. 

Wojtaszek \textit{et al.} \citep{Wojtaszek2014} used such a current-driven \textcolor{black}{DNP} scheme on a $^{13}$C graphene spin-valve device with ferromagnetic contacts at 4.2 K and low magnetic fields. However, the authors could not generate a sufficient nuclear Overhauser field that was resolvable in their experiments. Here, we test a different route to find signatures of the hyperfine interaction in $^{13}$C graphene.

We follow up on ealier attempts to study hyperfine effects by using laterally large multilayer $^{13}$C graphene (sample B) to increase the limit of detection\cite{Dora2010}. We remind the reader that graphene synthesized by MBE or CVD is comprised of grains\cite{Wofford2014} and that the fabrication process introduces \textcolor{black}{inhomogeneously } distributed \textsl{curvature} as existing in a large number of folds, bends and ripples. In carbon nanotubes, curvature induces $sp$-hybridization of the electron orbitals that affects the spin-orbit coupling\cite{Hernando2006, Kuemmeth2008} and the hyperfine interaction. The  \textcolor{black}{magnitude} of the hyperfine interaction in carbon nanostructures and how it is influenced by the curvature is still a controversial topic\cite{Pennington1996,Yazyev2008,Csiszar2014,Fischer2009,Pei2017}. In our sample, unintentional curvature exists as randomly distributed folds of various sizes and nanoripples and might result in a slightly larger overall hyperfine interaction than anticipated for more pure (exfoliated) graphene.

\begin{figure}[!]
   \includegraphics[width=0.45\textwidth, angle=0]{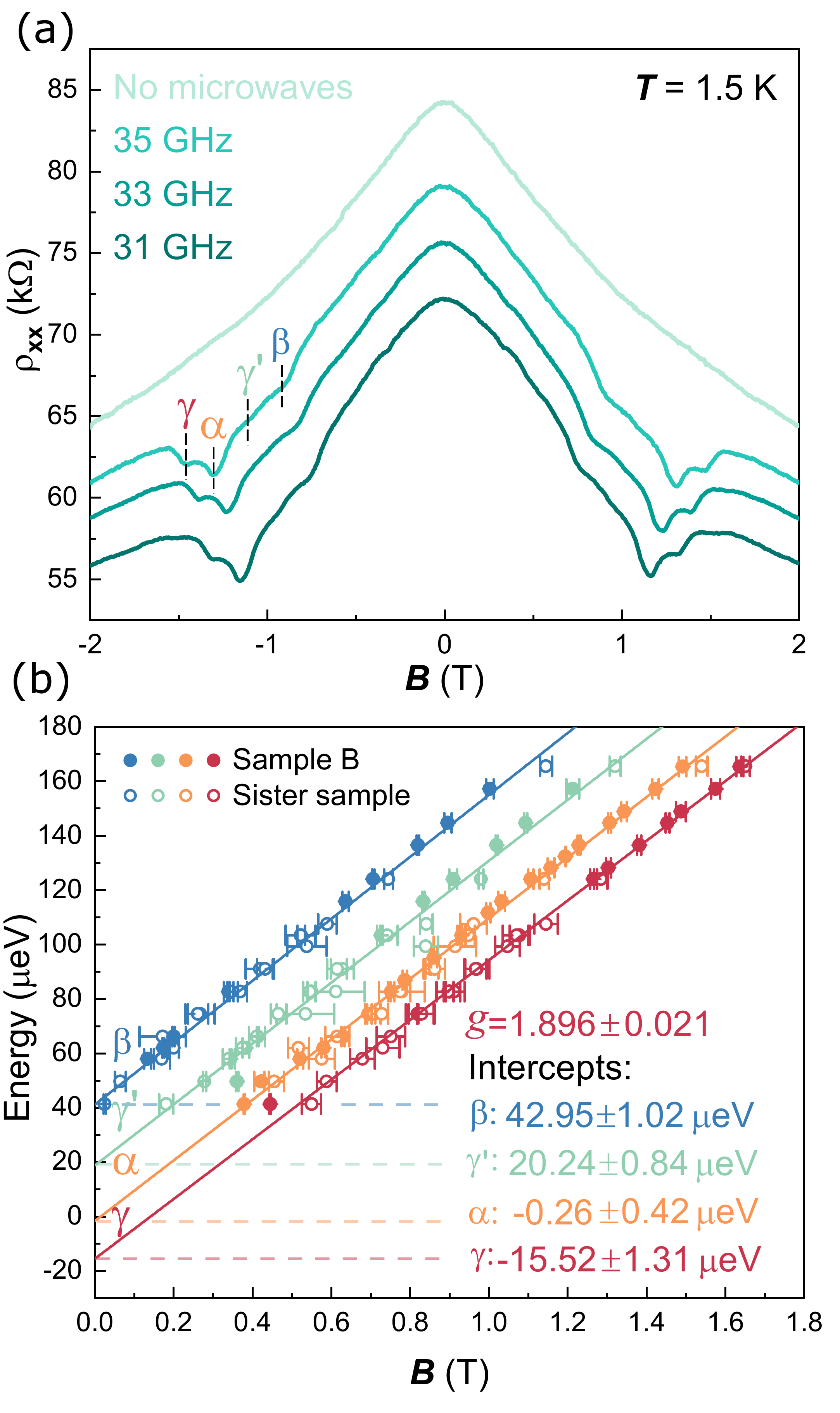}
   \caption{\textcolor{black}{(a) Comparison of $\rho_\mathrm{xx}$ without microwave (light green solid line) and with microwaves of 35 GHz, 33 GHz and 31 GHz all 22 dBm (shades of darker green lines)}, measured in few-layer $^{13}$C-graphene (sample B) at 1.5 K and with the gate voltage tuned close to the CNP. Resonant excitations are labeled by Greek letters $\alpha$, $\beta$, $\gamma$ (see main text for details). (b) Analysis of all resonances \textcolor{black}{[including those from an identical sister sample from the same processing batch]} shows the linear energy dispersion with magnetic field. The deduced  $g$-factor from the slope is \textcolor{black}{$g=1.896\pm0.021$}. The intercepts are \textcolor{black}{(42.95$\pm$1.02)$\mu$eV ($\beta$), (20.24$\pm$0.84)$\mu$eV ($\gamma$') and (-15.52$\pm$1.31)$\mu$eV ($\gamma$). The linear fit was not forced through the origin resulting in a tiny offset for the pure Zeeman energy represented by $\alpha$.}} 
  \label{fig4}
\end{figure}

Our second experimental adaption is the method that transfers momentum to the nuclei and probes the hyperfine interaction. Groundbreaking ESR experiments on GaAs two-dimensional electron systems had already shown the sensitivity of the ESR resonance to nuclear fields\cite{Berg1990}. Shchepetilnikov \textit{et al.}\cite{Shchepetilnikov2016} have more recently demonstrated the detection of nuclear fields using resistively-detected electron spin resonance (ESR) experiments on a AlAs 2D electron system. AlAs is a semiconductor with a comparable electron $g$-factor and $p$-type orbitals of the conduction electrons. Hence, the $p$-orbital character in graphene should not preclude a similar experimental approach.

The resistively detected ESR method is based on monitoring the sample resistance in a ramping magnetic field, $B$, under continuous irradiation with microwaves of constant power and frequency, $\nu$. Once the varying Zeeman splitting energy $g\cdot \mu_B\cdot B$ matches the (constant) energy of the microwave quanta, h$\nu$, electron spins are resonantly excited, a process which is signaled by a small change in the resistance. Here, $g$ is the electron $g$-factor. A nuclear magnetic field acts on the electron Zeeman energy and shifts the ESR transitions by $\Delta B \propto A_\mathrm{hf}/\mu_B g$. 

The low energy electronic band structure of graphene can be modified by intrinsic spin orbit splitting \cite{Mani2012, Sichau2019} and sublattice splitting  \cite{Singh2020}, resulting in multiple resonant excitations between bands of opposing spin and chirality, however. Figure \ref{fig4}(a) contrasts an exemplary magneto-resistance measurement in sample B with and without applied radiation of 35 GHz at 22 dBm at nominal $T$ = 1.5 K. Here, the density was tuned close to the CNP.

Under constant microwave radiation (orange solid line) multiple resonances are visible. We marked the resonances with Greek letters, using $\alpha$ for excitations between the pure Zeeman levels and $\beta$ for excitations between bands that are split by intrinsic spin-orbit coupling\cite{Sichau2019}. The two $\gamma$ lines result from sublattice splitting\cite{Singh2020}. Sublattice splitting can originate from a symmetry breaking due to the substrate but also from a (local) Bernal stacking in graphene multilayers \cite{Kindermann2011}. Following a previous report\cite{Mani2012}, we have fitted each resonance with a Lorenzian function to analyze the amplitudes and (center) magnetic field occurrences. 

\begin{figure}[!]
   \includegraphics[width=0.6\textwidth, angle=0]{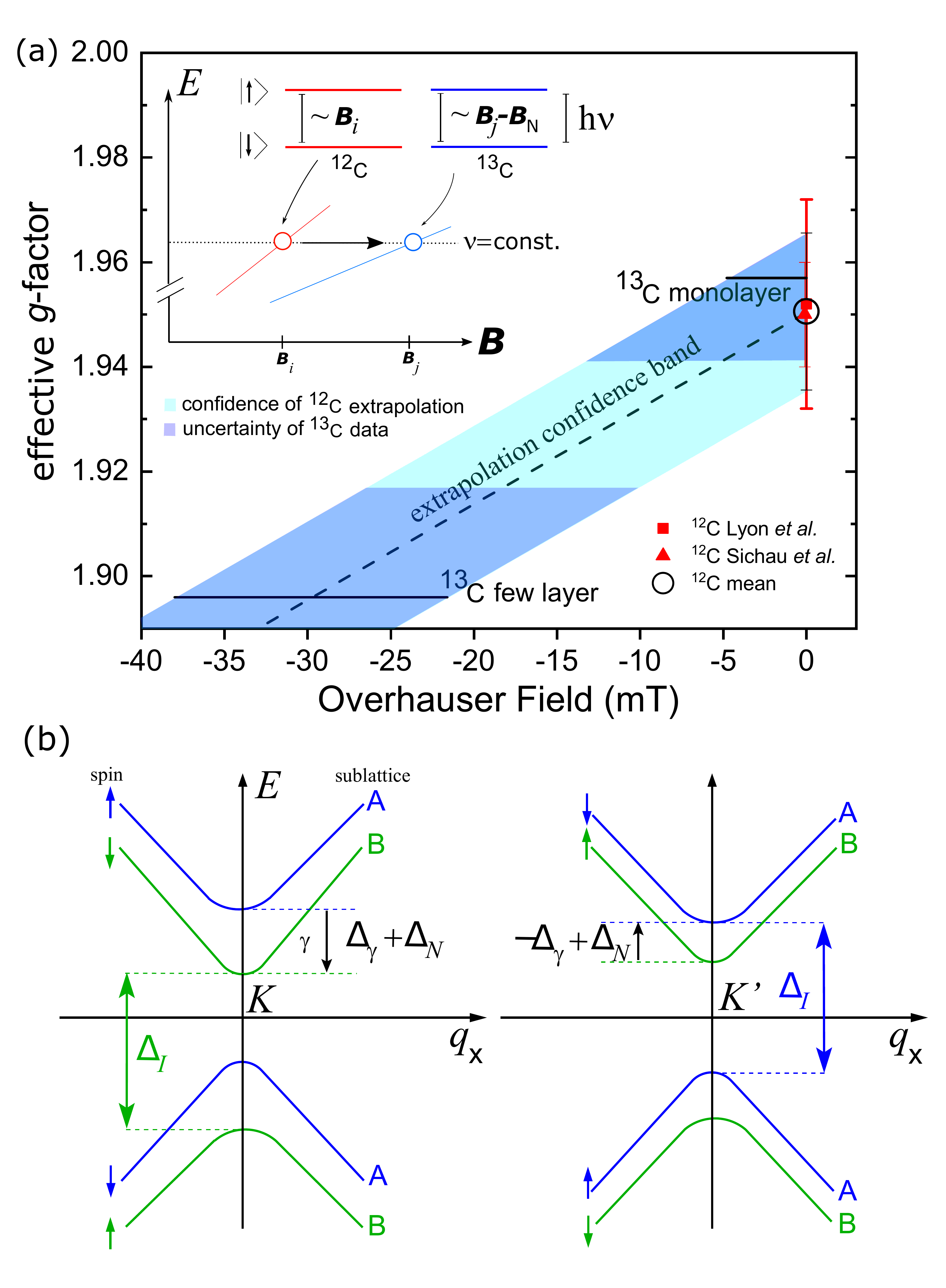}
   \caption{(a) Extrapolation of the $^{12}$C $g$-factor\cite{Lyon2017b, Sichau2019} in    the presence of an Overhauser field \textcolor{black}{(dashed line). The light blue band of confidence includes the measurement uncertainty.} The $g$-factors deduced from few-layer and monolayer $^{13}$C graphene are shown by the \textcolor{black}{black solid horizontal lines intersecting the extrapolation confidence band (dark shaded areas indicating data uncertainty).} The inset is a qualitative energy dispersion of the electron Zeeman splitting with a nuclear field $B_N<0$ (blue) and $B_N=0$ (red). A nuclear field will reduce the electron Zeeman splitting, i.e., for a constant $\nu$ a larger $B$ is needed to match the resonance condition. \textcolor{black}{Band diagram at $K$ and $K$' in the presence of spin and sublattice splitting\cite{Singh2020}, modified by a shift in energy of $\Delta_N$ induced by a nuclear field.}}
  \label{fig5}
\end{figure}

Figure \ref{fig4}(b) is a summary of this analysis, showing the linear dispersion of microwave energy $E$ = h$\nu$ versus the (center) magnetic field occurrences of all resonances measured at nominally 1.5 K. \textcolor{black}{Here, we have increased the number of data points by adding measurements from a sister few-layer $^{13}$C sample that is identical to sample B and from the same processing batch. We have also} repeated measurements at much higher carrier density and obtained the same value, \textcolor{black}{confirming that Rashba spin-orbit interaction is negligible. However, other types of spin-orbit interactions, arising from a hBN substrate for example, can modify the effective $g$-factor.\cite{Singh2020} We excluded these effects by exclusively using similar SiO$_2$ substrates for all samples [and we used the same fabrication methods to process the graphene].}

\textcolor{black}{The data in Fig. \ref{fig4}(b) exhibit two signatures that are consistent with the presence of a small nuclear field. a.} From a linear regression of the data points of the $\alpha$ resonances that directly follow h$\nu=g\cdot \mu_B\cdot B$, we deduced the effective electron $g$-factor of \textcolor{black}{$g=1.896 \pm 0.021$}, which is about \textcolor{black}{(2.95$\pm$1.16)}$\%$ smaller than determined from ESR measurements on $^{12}$C graphene on a SiO$_2$ or SiC substrate. \cite{Mani2012, Lyon2017b, Sichau2019, Anlauf2021, Prada2021} \textcolor{black}{b. We observe an asymmetry that represents the energy splitting of sublattice and spin degrees of freedom.}

\textcolor{black}{We first address the reduction of the $g$-factor, and we will 
begin by excluding various effects. i.} Unlike the \underline{nuclear} Zeeman splitting that is probed in nuclear magnetic resonance experiments, the \underline{electron} Zeeman energy, or more precisely $\mu_B$, does not depend on the isotope \textcolor{black}{mass. ii. Since the $g$-factor of bulk graphite is significantly larger and exceeds a value of 2, a possible layer-dependence should lead to higher but not smaller values. It was also demonstrated that even the $g$-factor of trilayer graphene is still independent on the number of layers\cite{Mani2012} and still smaller than for bulk graphite. iii. There exists no dependence on carrier concentration and carrier type\cite{Lyon2017b}. iv. We had excluded possible effects arising from varying substrates or a varying sample preparation. The reduction of the $g$-factor, however, is consistent with the presence of a nuclear field since a sizable nuclear field will affect the electron Zeeman splitting\cite{Berg1990, Kamp1992, Hillman2001, Shchepetilnikov2016}, which will be reflected in the experimentally deduced (i.e., effective) $g$-factor.} The small variance \textcolor{black}{between the electron $g$-factors for $^{12}$C\cite{Lyon2017b} and $^{13}$C graphene which is of the order of $\Delta g \approx$ 0.054$\pm$0.023} may thus signal a small nuclear field.

The nuclear field can be expressed as\cite{Paget1977}

\begin{equation}
\label{eqn:nfield}
B_\mathrm{N} = f\cdot b_\mathrm{N}\frac{B\cdot \langle S\rangle B}{B^2},
\end{equation}

where $B$ is the external magnetic field, $\langle S\rangle$ represents the average electron spin polarization, $f$ a spin relaxation factor and $b_\mathrm{N}$ the effective magnetic field as a result of nuclear spin polarization. $b_\mathrm{N}$ has been estimated to be -5.2 mT for $^{13}$C graphene\citep{Wojtaszek2014}. We cannot calculate $B_\mathrm{N}$ because unambiguous absolute values for $f$ and $\langle S\rangle$ are not available. We, however, can estimate the hypothetical nuclear field that results in a certain effective $g$-factor.

A nuclear field will slightly reduce the electron Zeeman splitting. Since the microwave frequency is constant, the resonance condition for electron spin flips now requires a slightly larger external magnetic field, $B$, as illustrated in \textcolor{black}{inset of} Fig. \ref{fig5} (a). The larger magnetic field converts to a smaller effective $g$-factor. \textcolor{black}{The main panel of Fig. \ref{fig5} (a) shows an extrapolation of the experimental value of the $^{12}$C graphene $g$-factor (dashed line) and their uncertainties (light blue band of confidence). The extrapolation assumes a nuclear Overhauser field that acts on the Zeeman energy and thus the $g$-factor. Figure \ref{fig5} (a)} estimates that the \textcolor{black}{$\Delta g \approx$ 0.054} between $^{12}$C and few-layer $^{13}$C \textcolor{black}{(lower solid black horizontal line)} would then be consistent with an Overhauser field of the order of \textcolor{black}{(-29.8 mT $\pm$ 8.3) mT (lower black horizontal line)}. We also obtained an experimental $g$-factor of 1.957 $\pm$ 0.016 from sample A [single layer $^{13}$C graphene] \textcolor{black}{(upper horizontal line)}. This value is close to those for $^{12}$C graphene, but the large uncertainty still allows for a possible Overhauser field smaller than -5 mT. A layer effect on the $g$-factor was ruled out in earlier studies\cite{Mani2012} but for $^{13}$C graphene multilayers may simply be required to achieve nuclear fields. 

\textcolor{black}{We now focus on the asymmetry in the energy that represent the splitting of sublattice and spin degrees of freedom. Close inspection of the intercepts of the $\gamma$ and $\gamma$' lines with the energy axis at $B$ = 0 T in Fig. \ref{fig4}(b) reveal a significant asymmetry. The higher $\gamma$' reveals a zero-field splitting of about 20.2 $\mu$eV, whereas the lower $\gamma$ line indicates a gap of -15.5 $\mu$eV. This asymmetry is consistent with nuclear field that \textcolor{black}{increases} the gap at the $K$ point and \textcolor{black}{lowers} it at the $K$' point as shown in Fig. \ref{fig5} (c). \textcolor{black}{Note that the gap has a} different sign (and magnitude) at each  \textcolor{black}{DP}, owing to \textcolor{black}{the non-trivial band topology\cite{kane2005}, where spin and sublattice spin degrees of freedom are related by chirality. This results in spin-polarized} subbands with opposite ordering at the two $K$-points\cite{Singh2020}, causing the depicted band inversion (blue and green bands that represent the sublattice degree of freedom). 
Solving the zero-field splitting gaps $\Delta_\gamma+\Delta_N\approx(20.24 \pm 0.84)\mu$eV and $\Delta_\gamma-\Delta_N\approx(-15.52 \pm 1.31)\mu$eV with $\Delta_\gamma = (17.88 \pm 1.1)\mu$eV, we obtain an effective $|\Delta_N|\approx (2.36 \pm 0.24)\mu$eV which converts to a nuclear field of $|\Delta_N/g\mu_B|\approx(21.5 \pm 2.2)$ mT.}

\textcolor{black}{These two} estimated values of an Overhauser field are consistent with theoretical predictions and experimental studies\cite{Churchill2009, Palyi2009, Dora2010, Wojtaszek2014,Pei2017} and might have benefited from the aforementioned occurrences of local curvature in our large-area MBE graphene. We concede, however, that the variations are very small and graze the uncertainty of our measurements. 

We tried to prepare different nuclear polarizations by varying the microwave power and magnetic field sweep rate as well as reversing the magnetic field sweep direction. A reduction of the microwave power makes the resonances quickly vanish in the resistive background and fitting of the ESR line shapes becomes impossible. Changes in the magnetic field sweep rate or sweep direction did not exhibit any hysteretic effects or additional shifts that we could resolve within our existing error. DNP through ESR in $^{13}$C CVD graphene appears to work only while we resonantly excite electron spins. We assume that the smallness of the hyperfine interaction and the achievable nuclear field always results in an Overhauser field close to its maximal value when we perform ESR.

The ESR line shape would reflect the build-up of an Overhauser field\cite{Shchepetilnikov2016}. In our case, the line shape is always very broad while any hyperfine field will be very small. From the line shape we can only reliably deduce the spin diffusion times, $\tau_\mathrm{S}$, \textcolor{black}{which will be sensitive to spin-flip scattering.} The line width $\Delta B_{\mathrm{res}}$  is inversely proportional to the spin relaxation time\cite{Mani2012},

\begin{equation}
     \tau_\mathrm{S} = \frac{h}{4\pi \Delta E_{\mathrm{res}}} = \frac{h}{4\pi\cdot g\cdot \mu_B\cdot\Delta B_\mathrm{res}}.
\end{equation}

Spin diffusion times were reported to be mostly independent of the carbon isotope and ranged between 60 and 80 ps and with weak dependence on the carrier concentration\cite{Wojtaszek2014}. From a Lorentzian fit of our resonances close to the CNP we obtained the spin diffusion times $\tau_\mathrm{S}^{12\mathrm{C}} \approx (70 \pm 37)$ ps \textcolor{black}{(the $^{12}$C reference sample used in section \ref{sec:wal})}, $\tau_\mathrm{S}^{13\mathrm{C, B}} \approx (12 \pm 9)$ ps \textcolor{black}{(sample B)} and \textcolor{black}{$\tau_\mathrm{S}^{13\mathrm{C, A}}\approx (9 \pm 8)$ ps} \textcolor{black}{(sample A). The near parity of $\tau_\mathrm{S}$ for sample A and B implies that the D-line in the Raman measurement at room temperature is not relevant for low temperature electron spin resonance.} 

For completeness, we conclude by noting that we also exposed the sample to radio frequencies that match the nuclear spin splitting (NMR) of the $^{13}$C nuclei after performing DNP via ESR. However, with our currently available measurement scheme that limits ESR to magnetic fields smaller than 1.5 Tesla (and 1.5 K), we were unable to detect significant and \textcolor{black}{reproducible} signals in the resistive background.

\section{CONCLUSION}

In this study, we have revisited the important question of whether the hyperfine interaction between the spins of electrons and nuclei can experimentally be addressed in isotopically purified $^{13}$C graphene. \textcolor{black}{Our study confirms the elusive and very weak character of hyperfine interactions found in earlier studies, but we highlight two possible routes to access this rich regime. An analysis of the \textcolor{black}{WL} phenomenon in single layer $^{12}$C and $^{13}$C graphene shows an anomaly in the scattering times that exhibit a strong suppression around the CNP for $^{13}$C. In the framework of the existing \textcolor{black}{WL} spinless model, nuclear spins appear to act as strongly dephasive and atomically sharp scattering centers that are not sufficiently screened at low carrier concentrations.} In multilayer $^{13}$C graphene, we used microwave activated transport experiments to resonantly excite electron spins and \textcolor{black}{search for signatures of} the hyperfine interaction with the nuclei. \textcolor{black}{We found a reduced effective $g$-factor that hints the presence of a small nuclear magnetic field and its impact on the electron Zeeman energy. We also observed an asymmetry (shift) in the two resonances that represent the sublattice splitting. \textcolor{black}{The corresponding zero-field splitting} is also consistent with a nuclear field of about the same size as the one we deduced from the $g$-factor.}


\begin{acknowledgments}
We would like to thank Guido Burkard for insightful discussions. We wish to thank the Deutsche Forschungsgemeinschaft (DFG) with the Grant No.  BL-487/14-1 for its support. The authors also would like to acknowledge the Partnership for Innovation, Education, and Research (PIER). All measurements were performed with $^\mathrm{nano}$meas\cite{nanomeas}.
\end{acknowledgments}



\nocite{*}
\bibliography{strenzke_manuscript_resub.bbl}
\bibliographystyle{ieeetr}

\end{document}